\documentclass[twocolumn]{aastex63}

\shorttitle{Backyard Worlds T Subdwarfs}
\shortauthors{Meisner et al.}

\begin{document}

\title{New Candidate Extreme T Subdwarfs from the Backyard Worlds: Planet 9 Citizen Science Project}

\correspondingauthor{Aaron M. Meisner}
\email{aaron.meisner@noirlab.edu}

\author[0000-0002-1125-7384]{Aaron M. Meisner}
\affiliation{NSF's National Optical-Infrared Astronomy Research Laboratory, 950 N. Cherry Ave., Tucson, AZ 85719, USA}

\author[0000-0002-6294-5937]{Adam C. Schneider}
\affil{United States Naval Observatory, Flagstaff Station, 10391 West Naval Observatory Rd., Flagstaff, AZ 86005, USA}
\affil{Department of Physics and Astronomy, George Mason University, MS3F3, 4400 University Drive, Fairfax, VA 22030, USA}

\author[0000-0002-6523-9536]{Adam J. Burgasser}
\affiliation{Center for Astrophysics and Space Science, University of California San Diego, La Jolla, CA 92093, USA}

\author[0000-0001-7519-1700]{Federico Marocco}
\affiliation{IPAC, Mail Code 100-22, California Institute of Technology, 1200 E. California Blvd., Pasadena, CA 91125, USA}

\author[0000-0002-2338-476X]{Michael R. Line}
\affiliation{School of Earth \& Space Exploration, Arizona State University, Tempe AZ 85287, USA}

\author[0000-0001-6251-0573]{Jacqueline K. Faherty}
\affiliation{Department of Astrophysics, American Museum of Natural History, Central Park West at 79th Street, New York, NY 10024, USA}

\author[0000-0003-4269-260X]{J. Davy Kirkpatrick}
\affiliation{IPAC, Mail Code 100-22, California Institute of Technology, 1200 E. California Blvd., Pasadena, CA 91125, USA}

\author[0000-0001-7896-5791]{Dan Caselden}
\affiliation{Gigamon Applied Threat Research, 619 Western Ave., Suite 200, Seattle, WA 98104, USA}

\author[0000-0002-2387-5489]{Marc J. Kuchner}
\affiliation{NASA Goddard Space Flight Center, Exoplanets and Stellar Astrophysics Laboratory, Code 667, Greenbelt, MD 20771, USA}

\author{Christopher R. Gelino}
\affiliation{IPAC, Mail Code 100-22, California Institute of Technology, 1200 E. California Blvd., Pasadena, CA 91125, USA}

\author[0000-0002-2592-9612]{Jonathan Gagn\'e}
\affiliation{Institute for Research on Exoplanets, Universit\'e de Montr\'eal, 2900 Boulevard \'Edouard-Montpetit Montr\'eal, QC Canada H3T 1J4}
\affiliation{Plan\'etarium Rio Tinto Alcan, Espace pour la Vie, 4801 av. Pierre-de Coubertin, Montr\'eal, Qu\'ebec, Canada}

\author[0000-0002-9807-5435]{Christopher Theissen}
\altaffiliation{NASA Sagan Fellow}
\affiliation{Center for Astrophysics and Space Science, University of California San Diego, La Jolla, CA 92093, USA}

\author{Roman Gerasimov}
\affiliation{Center for Astrophysics and Space Science, University of California San Diego, La Jolla, CA 92093, USA}

\author{Christian Aganze}
\affiliation{Center for Astrophysics and Space Science, University of California San Diego, La Jolla, CA 92093, USA}

\author[0000-0002-5370-7494]{Chih-chun Hsu}
\affiliation{Center for Astrophysics and Space Science, University of California San Diego, La Jolla, CA 92093, USA}

\author[0000-0001-9209-1808]{John P. Wisniewski}
\affiliation{Homer L. Dodge Department of Physics and Astronomy, University of Oklahoma, 440 W. Brooks Street, Norman, OK 73019, USA}

\author[0000-0003-2478-0120]{Sarah L. Casewell}
\affiliation{Department of Physics and Astronomy, University of Leicester, University Road, Leicester LE1 7RH, UK}

\author[0000-0001-8170-7072]{Daniella C. Bardalez Gagliuffi}
\affiliation{Department of Astrophysics, American Museum of Natural History, Central Park West at 79th Street, New York, NY 10024, USA}

\author[0000-0002-9632-9382]{Sarah E. Logsdon}
\affiliation{NSF's National Optical-Infrared Astronomy Research Laboratory, 950 N. Cherry Ave., Tucson, AZ 85719, USA}

\author{Peter R. M. Eisenhardt}
\affiliation{Jet Propulsion Laboratory, California Institute of Technology, 4800 Oak Grove Drive, M/S 169-327, Pasadena, CA 91109, USA}

\author[0000-0003-0580-7244]{Katelyn Allers}
\affiliation{Physics and Astronomy Department, Bucknell University, 701 Moore Ave., Lewisburg, PA 17837, USA}

\author[0000-0002-1783-8817]{John H. Debes}
\affiliation{ESA for AURA, Space Telescope Science Institute, 3700 San Martin Drive, Baltimore, MD 21218, USA}

\author{Michaela B. Allen}
\affiliation{NASA Goddard Space Flight Center, Exoplanets and Stellar Astrophysics Laboratory, Code 667, Greenbelt, MD 20771, USA}

\author{Nikolaj Stevnbak Andersen}
\affiliation{Backyard Worlds: Planet 9}

\author[0000-0003-2236-2320]{Sam Goodman}
\affiliation{Backyard Worlds: Planet 9}

\author[0000-0002-8960-4964]{L\'eopold Gramaize}
\affiliation{Backyard Worlds: Planet 9}

\author{David W. Martin}
\affiliation{Backyard Worlds: Planet 9}

\author[0000-0003-4864-5484]{Arttu Sainio}
\affiliation{Backyard Worlds: Planet 9}

\author[0000-0001-7780-3352]{Michael C. Cushing}
\affiliation{Department of Physics and Astronomy, University of Toledo, 2801 West Bancroft St., Toledo, OH 43606, USA}

\author{The Backyard Worlds: Planet 9 Collaboration}

\begin{abstract}

\cite{esdTs} presented the discovery of WISEA J041451.67$-$585456.7 and WISEA J181006.18$-$101000.5, which appear to be the first examples of extreme T-type subdwarfs (esdTs; metallicity $\le -1$ dex, $T_{\rm eff} \lesssim 1400$ K). Here we present new discoveries and follow-up of three T-type subdwarf candidates, with an eye toward expanding the sample of such objects with very low metallicity and extraordinarily high kinematics, properties that suggest membership in the Galactic halo. Keck/NIRES near-infrared spectroscopy of WISEA J155349.96+693355.2, a fast-moving object discovered by the Backyard Worlds: Planet 9 citizen science project, confirms that it is a mid-T subdwarf. With $H_{W2} = 22.3$ mag, WISEA J155349.96+693355.2 has the largest W2 reduced proper motion among all spectroscopically confirmed L and T  subdwarfs, suggesting that it may be kinematically extreme. Nevertheless, our modeling of the WISEA J155349.96+693355.2 near-infrared spectrum indicates that its metallicity is only mildly subsolar. In analyzing the J155349.96+693355.2 spectrum, we present a new grid of low-temperature, low-metallicity model atmosphere spectra. We also present the discoveries of two new esdT candidates, CWISE J073844.52$-$664334.6 and CWISE J221706.28$-$145437.6, based on their large motions and colors similar to those of the two known esdT objects. Finding more esdT examples is a critical step toward mapping out the spectral sequence and  observational properties of this newly identified population.

\end{abstract}


\section{Introduction}
\label{sec:intro}

What are the physical properties of the lowest luminosity substellar objects in our Milky Way Galaxy? How do brown dwarf atmospheres and spectral energy distributions change as a function of multiple variables including temperature, metallicity, and age? How has the birth rate of substellar objects evolved over time, from early periods of star formation in the Universe to the present epoch? Ultracool subdwarfs, very low mass stars and brown dwarfs with low metallicity which can be found in the Milky Way thick disk or halo, are important laboratories that play a major role in answering these questions \citep[e.g.,][]{burgasser_subdwarf_overview}.

The solar neighborhood provides our best opportunity to study the Galactic substellar population across its span of temperatures, metallicities, ages, and masses. Pinpointing nearby substellar objects representing the extremes of these parameters is therefore a key way to test/develop theoretical models, such as those of giant exoplanet atmospheres \citep[e.g.,][]{bd_exo_connection, marley_review, leggett_white_paper} and the star formation process at low masses \citep[e.g.,][]{davy_white_paper, cw_byw_20pc}.

Hundreds of T dwarfs \citep[$450 \ \textrm{K} \ \lesssim T_{\rm eff} \lesssim 1400$ K; e.g.,][]{gl229b, kirkpatrick11, mace_t_dwarfs} and dozens of low-metallicity T subdwarfs \citep[e.g.,][]{burningham14,pinfield_methodology,zhang_sdt} are now known. Recently, the Backyard Worlds: Planet 9 citizen science project \citep[\href{http://backyardworlds.org}{backyardworlds.org};][henceforth Backyard Worlds]{backyard_worlds} discovered the first two examples of `extreme T-type subdwarfs' \citep[esdTs;][]{esdTs}, with $T_{\rm eff} \lesssim 1,400$ K and [Fe/H] $\le -1$. The esdTs have distinctive near-infrared (NIR) colors unlike those of other brown dwarfs and spectra not well-replicated by any existing models. Determining the space densities of very cold and metal-poor objects, such as esdTs, may yield new insights about the dependence of low mass star formation on metallicity throughout cosmic time. The low-metallicity atmospheres of the coldest isolated subdwarfs will also help us understand (by proxy) the atmospheres of giant exoplanets that orbit old and/or metal-poor host stars.

To further map out the parameter space of T-type subdwarfs with very low metallicity and extreme kinematics, we searched through thousands of Backyard Worlds moving object discoveries, identifying new esdT candidates based on peculiar infrared colors and large proper motion. This yielded three new extreme T subdwarf candidates: CWISE J073844.52$-$664334.6 (hereafter CWISE 0738$-$6643), WISEA J155349.96+693355.2 \citep[][hereafter WISEA 1553+6933]{byw_spitzer_phot} and CWISE J221706.28$-$145437.6 (hereafter CWISE 2217$-$1454).

In $\S$\ref{sec:byw_overview} we provide a brief overview of the Backyard Worlds citizen science project. In $\S$\ref{sec:target_selection} we describe our selection of new esdT candidates from within the full list of Backyard Worlds moving object discoveries. In $\S$\ref{sec:candidates} we introduce the three esdT candidates identified. In $\S$\ref{sec:spectro} we present Keck/NIRES spectroscopy of WISEA 1553+6933, including comparison against an extensive new grid of low-temperature, low-metallicity model atmosphere spectra. In $\S$\ref{sec:synth} we investigate the synthetic colors of our new models as a function of metallicity. We conclude in $\S$\ref{sec:conclusion}.

\section{The Backyard Worlds: Planet 9 Citizen Science Project}
\label{sec:byw_overview}

Backyard Worlds \citep{backyard_worlds} is a moving object search that crowdsources the visual inspection of all-sky \textit{Wide-field Infrared Survey Explorer} \citep[$WISE$;][]{wright10} images among thousands of online volunteers. The project launched in 2017 February via the Zooniverse web platform \citep{zooniverse}. Volunteers scrutinize  time series image blinks (referred to as `flipbooks'), with each flipbook showing the evolution of a random $10' \times 10'$ sky patch over the 2010-2016 time period. Each flipbook frame is a two-color composite representing one \textit{WISE} sky pass, with \textit{WISE} W1 (3.4~$\mu$m) encoded as the blue channel and W2 (4.6~$\mu$m) as the red channel. Backyard Worlds pushes fainter than prior \textit{WISE}-based motion searches because its flipbooks are built from deep/clean time-resolved `unWISE' coadds \citep{lang14, tr_neo2} that offer a long $\sim$6 year time baseline. The signature Backyard Worlds science application is discovering cold brown dwarfs in the solar neighborhood based on their characteristic combination of high proper motion and red W1$-$W2 color. Backyard Worlds discoveries have spanned multiple subtopics within solar neighborhood science, including extremely cold Y dwarfs \citep{w0830, byw_spitzer_phot}, co-moving companions \citep[e.g.,][]{wise2150, rothermich, jalowiczor}, T dwarfs \citep[e.g.,][]{backyard_worlds}, white dwarf disks \citep{j0207}, and previously overlooked members of the 20 pc brown dwarf census \citep{cw_byw_20pc}. Backyard Worlds revealed the first two known examples of extreme T subdwarfs \citep{esdTs}.

Backyard Worlds motion searches have expanded significantly beyond the project's Zooniverse interface. For instance, Backyard Worlds citizen scientists have created novel visualization tools to interactively customize \textit{WISE} image blinks \citep[e.g., WiseView;][]{wiseview}. A group of $\sim$300 advanced users has conducted numerous catalog-based brown dwarf candidate queries on platforms like IRSA  and Astro Data Lab \citep{data_lab_aspc}, particularly leveraging the unWISE Catalog \citep{unwise_catalog} and CatWISE 2020  \citep{catwise2020}, the latter offering \textit{WISE}-based proper motion measurements $>$10$\times$ more accurate than those of AllWISE \citep{cutri13}. Because Backyard Worlds participants scan \textit{WISE} images by eye for any/all moving sources, the project is well-positioned to discover atypical objects like subdwarfs that might be missed by traditional color searches \citep[e.g.,][]{griffith_brown_dwarfs}.

\section{Selection of New Extreme T Subdwarf Candidates}
\label{sec:target_selection}

As of 2021 April, Backyard Worlds participants have discovered roughly 3,200 motion-confirmed L, T, and Y dwarf candidates. Most of these discoveries are likely to be `ordinary' L and T dwarfs. We mined this large list of Backyard Worlds discoveries to determine whether any may be promising new candidate members of the esdT spectral class. Typically, very little information about our Backyard Worlds motion discoveries is available, since the sample is generally quite faint and red by selection. In many cases only W1 and W2 detections are available from archival survey data sets.

\begin{figure*}
\plotone{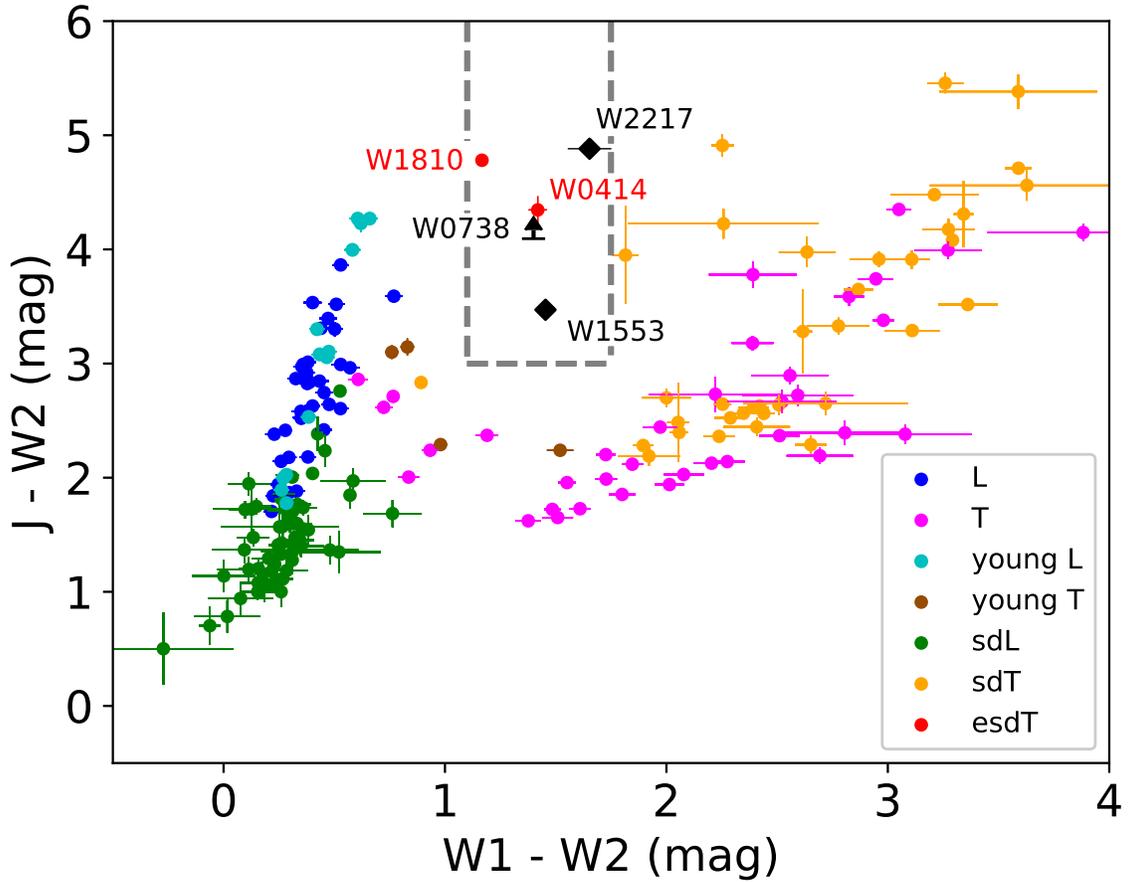}
\caption{J$-$W2 versus W1$-$W2 color-color plot showing the populations of L dwarfs (blue), T dwarfs (magenta), young L dwarfs within 20 pc \citep[cyan;][]{cw_byw_20pc}, young isolated T dwarfs (brown), L subdwarfs (green), T subdwarfs (orange), and extreme T subdwarfs (red). The two known esdT objects from \cite{esdTs} are individually labeled. WISEA 1553+6933 and CWISE 2217$-$1454 are labeled with black diamonds. CWISE 0738$-$6643, a new esdT candidate first presented in this work, is labeled with an upward arrow to indicate that its non-detection in VHS provides a lower limit of J$-$W2 = 4.09 mag. WISEA 1553+6933 and CWISE 2217$-$1454 inhabit unusual regions of this color-color space, suggesting that they could be either sdT or esdT. CWISE 0738$-$6643 seems to most closely align with the esdTs. Our adopted esdT candidate color-color selection box is shown as a gray dashed outline.}
\label{fig:color_color}
\end{figure*}

Aside from W1 and W2, J band is the most valuable filter at our disposal for which sensitive archival survey photometry exists over most of the sky. We therefore sought to identify esdT candidates based on the combination of W1, W2 and J photometry. esdTs inhabit a distinctive region of J$-$W2 versus W1$-$W2 color-color space (see Figure \ref{fig:color_color}, which is based on Figure 3 of \citealt{esdTs}). esdTs have much redder J$-$W2 colors than do T dwarfs in the same 1.1 mag $\le$ W1$-$W2  $\le$ 1.75 mag \textit{WISE} color range.

For all $\sim$3,200 Backyard Worlds moving object discoveries, we compiled available 2MASS \citep{tmass}, UHS\footnote{The UKIDSS project is defined in \cite{ukidss}. UKIDSS uses the UKIRT Wide Field Camera \citep[WFCAM;][]{wfcam} and a photometric system described in \cite{hewett2006}. The pipeline processing and science archive are described in \cite{irwin08} and \cite{hambly08}. We have used data from UHS DR1, which is described in detail in \cite{uhs}.} \citep{uhs}, and VHS \citep{vhs} J band photometry. We also took into consideration photometric follow-up that our team has obtained for a relatively small subset of the full Backyard Worlds discovery list. Lastly, when J band imaging is available from UHS or VHS but no J band counterpart is present, we compute 5$\sigma$ depth limits at the relevant sky location according to the procedure described in $\S$4.1 of \cite{esdTs}. This allows us to identify esdT candidates on the basis of large J$-$W2 color lower limits, even when a J detection is not available.

In our fiducial esdT candidate color-color selection box defined by J$-$W2 $ > $ 3 mag and 1.1 mag $<$ W1$-$W2 $<$ 1.75 mag (see Figure \ref{fig:color_color}), we found three new esdT candidates that remained viable after visual inspection to weed out problematic photometry (e.g., from \textit{WISE} blending). We also required a W1 SNR of at least 10 in order to ensure that the measured W1$-$W2 color is reliable. In $\S$\ref{sec:candidates} we discuss each of our three photometrically selected T subdwarf candidates in detail.

\section{T Subdwarf Candidates}
\label{sec:candidates}

\begin{figure*}
\plotone{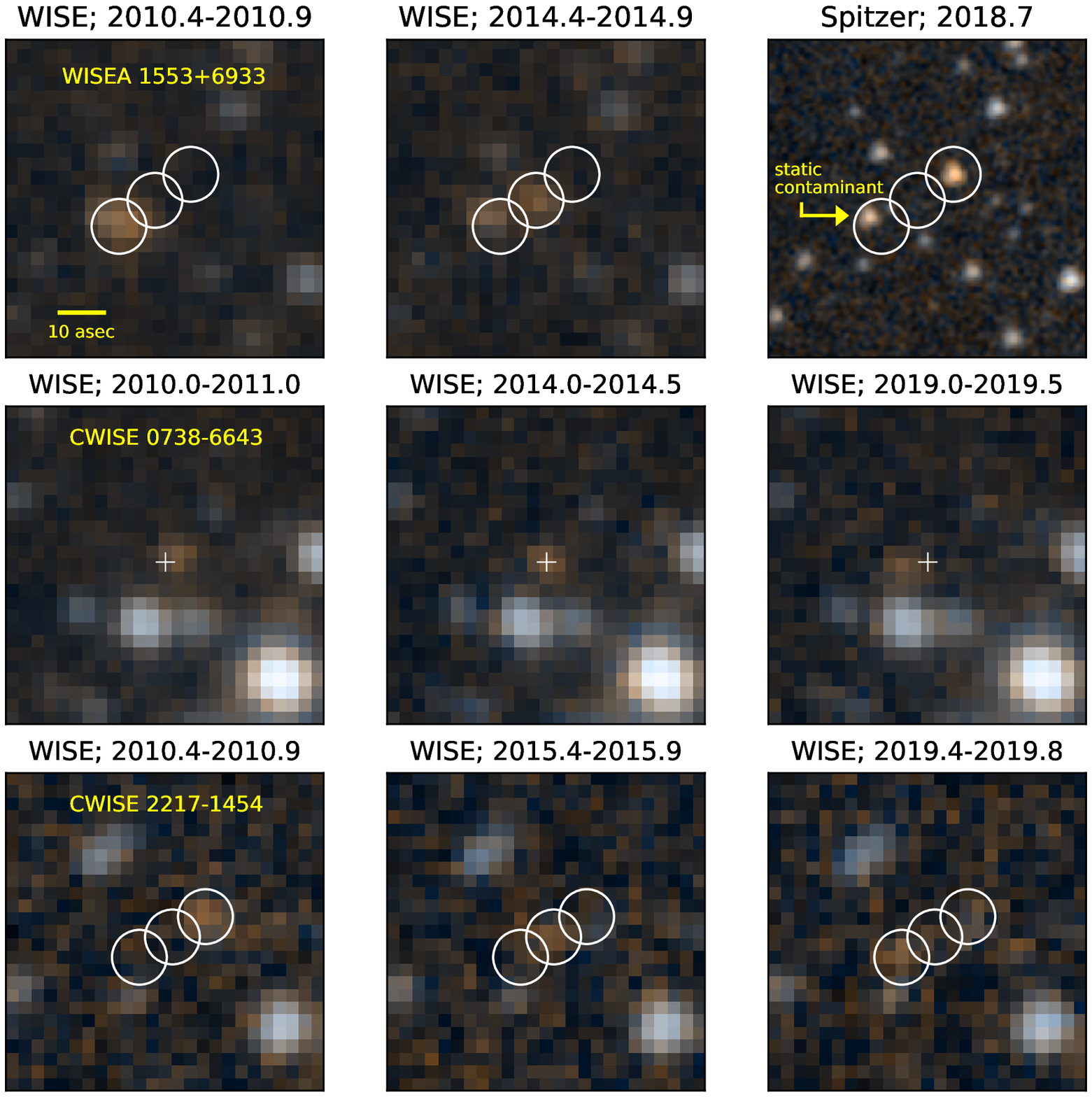}
\caption{Top row: Time series of \textit{WISE} and \textit{Spitzer} images illustrating the motion of WISEA 1553+6933. Each panel is a two-band color composite rendering. In each case W1 (ch1) is represented by the blue color channel and W2 (ch2) is represented by the red color channel. \textit{WISE} images are one-year coadds spanning calendar 2010 (left) and the first year of NEOWISE-Reactivation (center). The \textit{Spitzer} color composite at right is built from PID14076 IRAC imaging (PI: Faherty). East is left and north is up. The white circles track the northwesterly motion of WISEA 1553+6933 over the $\sim$2010-2019 time period. As seen in the left panel, WISEA 1553+6933 was severely blended with a static background object during 2010. Middle row: \textit{WISE} coadds illustrating the motion of CWISE 0738$-$6643. The white plus mark remains fixed to highlight the subdwarf candidate's southeasterly motion. Bottom row: \textit{WISE} coadds illustrating the motion of CWISE 2217$-$1454. All postage stamps are $1.1' \times 1.1'$ in angular extent. }
\label{fig:cutouts}
\end{figure*}

Figure \ref{fig:cutouts} illustrates the large proper motions ($\mu \sim 0.9-2.2''$/yr) of our three new subdwarf candidates as seen in \textit{WISE}. The following subsections provide additional details about each subdwarf candidate's photometry and kinematics.

\subsection{CWISE J073844.52$-$664334.6}
\label{sec:w0738}

CWISE 0738$-$6643 was discovered by Backyard Worlds participant L\'eopold Gramaize. CWISE 0738$-$6643 has W1$-$W2 = 1.40 $\pm$ 0.05 mag and is undetected in J band imaging from VHS, yielding a J$-$W2 $>$ 4.09 mag color limit. In the Figure \ref{fig:color_color} color-color plot, its J band lower limit places it at a location nearly identical to that of the esdT WISEA 0414$-$5854 \citep{esdTs}. CWISE 0738$-$6643 is at least 2.15 magnitudes redder in J$-$W2 than would be expected for a typical T dwarf of its same W1$-$W2 color. Table \ref{tab:phot_cand} provides photometric and astrometric information for CWISE 0738$-$6643. Using VHS imaging, we find a 5$\sigma$ limit of $K_S > 18.22$ mag, but this limit is not deep enough to meaningfully constrain the nature of CWISE 0738$-$6643.

\begin{deluxetable*}{cccccccc}
\tablecaption{Photometric subdwarf candidate discoveries}
\tablehead{
\colhead{CWISE Name} & W1 & W2 & $J_{MKO}$ & $K_S$ & $\mu_{\alpha}$ (mas/yr) & $\mu_{\delta}$ (mas/yr) & $\mu$ (mas/yr)
}
\startdata
\label{tab:phot_cand}
J073844.52$-$664334.6 & 17.221 $\pm$ 0.038 & 15.821 $\pm$ 0.036 & $>$ 19.92 & $>$ 18.22 & 765 $\pm$ 32 & $-432$
$\pm$ 30 & 878 $\pm$ 31 \\
J221706.28$-$145437.6 & 17.428 $\pm$ 0.078 & 15.775 $\pm$ 0.055 & 20.66 $\pm$ 0.02 & $>$ 18.20 & 1637 $\pm$ 65 & $-919$ $\pm$ 63 & 1878 $\pm$ 65 \\
\enddata
\tablecomments{New discoveries identified as extreme T subdwarf candidates based on their photometry and large motions. $WISE$ magnitudes (uncertainties) come from the w1mpro\_pm, w2mpro\_pm (w1sigmpro\_pm, w2sigmpro\_pm) columns of the CatWISE 2020 catalog \citep{catwise2020}. Proper motions are from CatWISE 2020. Quoted $\mu_{\alpha}$ values incorporate the cos($\delta$) factor. All magnitude limits are $5\sigma$. $K_{S}$ limits are based on archival VHS imaging.}
\end{deluxetable*}

\begin{figure}
\plotone{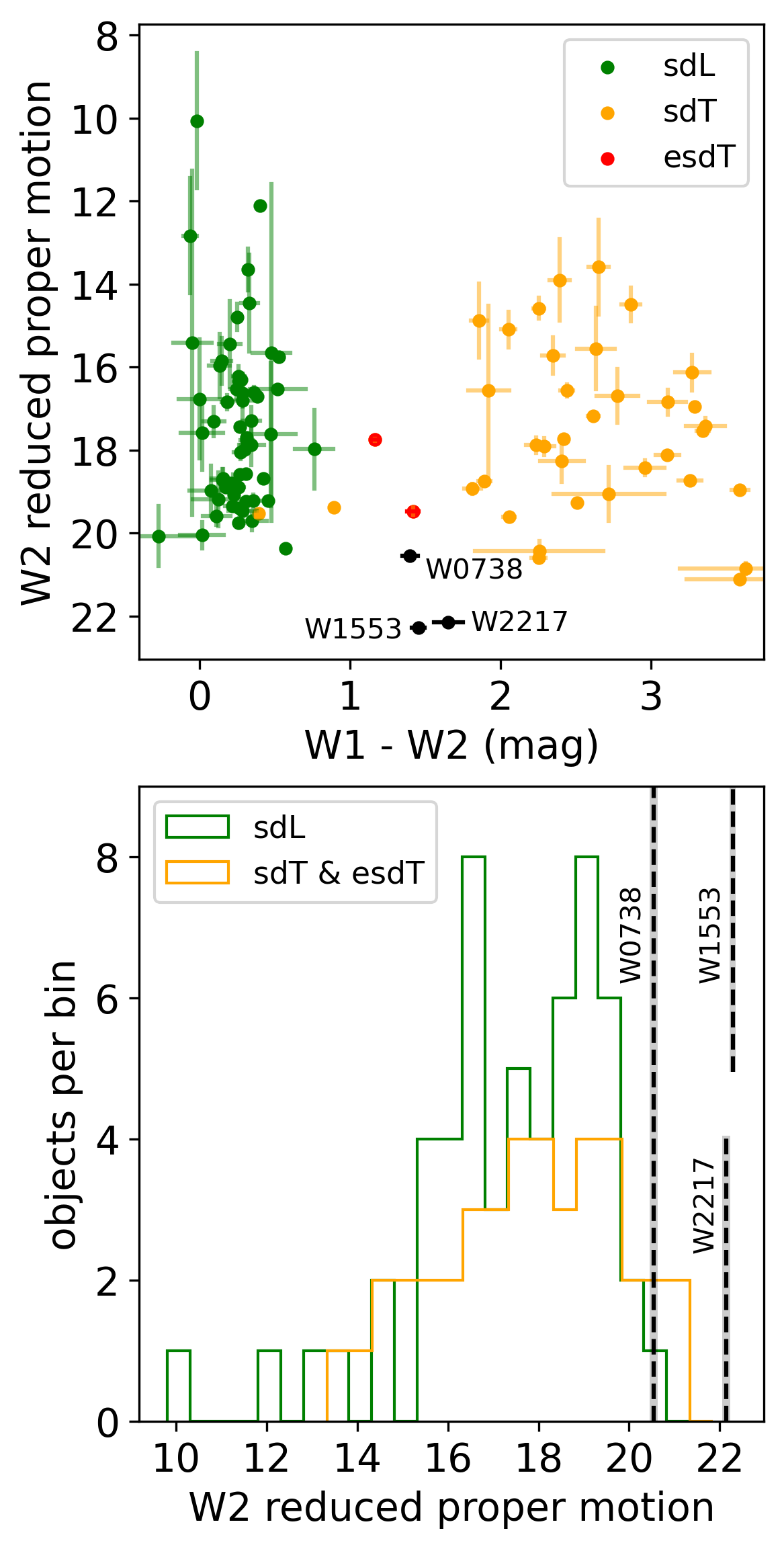}
\caption{W2 reduced proper motions $(H_{W2})$ for this work's sample in comparison to the population of spectroscopically confirmed sdL, sdT and esdT objects drawn from the prior literature. Top: $H_{W2}$ as a function of W1$-$W2 color. Bottom: Histograms of sdL and sdT $H_{W2}$ values with vertical dashed lines indicating the $H_{W2}$ values of WISEA 1553+6933, CWISE 2217$-$1454 and CWISE 0738$-$6643. WISEA 1553+6933 has the largest W2 reduced proper motion among all spectroscopically confirmed L and T subdwarfs.}
\label{fig:sd_rpm}
\end{figure}

In addition to falling within our esdT candidate color-color selection box, CWISE 0738$-$6643 has a large total proper motion of $\mu = 878 \pm 31$ mas/yr and a high W2 reduced proper motion\footnote{Defined as $H_{W2} = m_{W2} + 5$log$_{10}\mu + 5$, where $\mu$ is the total proper motion in arcseconds per year and $m_{W2}$ is the W2 apparent magnitude \citep{luyten22}.} ($H_{W2}$) similar to the largest $H_{W2}$ values among the prior literature's population of spectroscopically confirmed T subdwarfs (see Figure \ref{fig:sd_rpm}). The compilations of literature sdT and sdL objects we use throughout this work were initially curated as part of the \cite{esdTs} analysis, and are largely based on lists from \cite{zhang_sdt} and \cite{zhang_sdl}.

As none of our subdwarf candidates in this work have trigonometric parallax measurements available, reduced proper motion is a particularly valuable indicator pointing toward high kinematics and/or low luminosity. The reduced proper motion formula is a variant of the equation defining absolute magnitude, where the parallax has been replaced by total proper motion, in essence using large motion as a proxy for nearness. Because reduced proper motion grows with both apparent magnitude and total proper motion, objects with very low luminosity and/or high tangential velocity tend to have distinctively large reduced proper motions. Without trigonometric parallaxes for our subdwarf candidates, we cannot determine the extent to which their very high reduced proper motions are attributable to low luminosity versus high kinematics.

Using the \cite{cw_byw_20pc} polynomial relations fit to the bulk T dwarf population, we can derive \textit{WISE}-based phototype and distance estimates for CWISE 0738$-$6643. From the W1$-$W2 color, we obtain a spectral type estimate of T5 $\pm$ 1.2, and a photometric distance estimate of $43.6^{+10.4}_{-8.4}$ pc (based on $M_{W2}$ inferred from the $WISE$ color). This photometric distance yields a $V_{tan}$ estimate of $181^{+44}_{-36}$ km/s. The large uncertainties on these values are primarily driven by the 0.46 mag scatter associated with the conversion from W1$-$W2 color to absolute magnitude \citep{cw_byw_20pc}. If CWISE 0738$-$6643 is sub-luminous at W2 relative to typical T dwarfs of the same W1$-$W2 color, its distance and tangential velocity will both be smaller than the estimates reported here\footnote{The luminosity sequence of T subdwarfs is presently not well-mapped. L subdwarfs do exhibit different luminosity versus type trends than field L dwarfs in certain photometric bands \citep[e.g.,][]{gonzales18}.}.

The CWISE 0738$-$6643 total proper motion value and its uncertainty from Table \ref{tab:phot_cand} both impact our $V_{tan}$ estimate. Thanks to the fact that CatWISE 2020 spans a long $\sim$9 year time baseline and incorporates 6 years of \textit{WISE}/NEOWISE imaging, it delivers a high S/N proper motion measurement despite the faintness of CWISE 0738$-$6643 and the low \textit{WISE} angular resolution (FWHM $\approx$ 6.5$''$). \cite{catwise2020} compared the CatWISE 2020 proper motion measurements against those of \textit{Gaia} DR2 \citep{gaia_dr2}, to assess the CatWISE 2020 proper motion accuracy and quoted uncertainties. CWISE 0738$-$6643 is at very high absolute ecliptic latitude ($\beta = -80.3^{\circ}$), meaning that Figure 23 of \cite{catwise2020}, which studies a region surrounding the south ecliptic pole, is the most applicable \textit{Gaia}-CatWISE comparison. Figure 23 (bottom right panel) of \cite{catwise2020} indicates that, near the south ecliptic pole, the median $\chi^2$ of the CatWISE 2020 versus \textit{Gaia} DR2 proper motion comparison is $\sim$2 for both $\mu_{\alpha}$ and $\mu_{\delta}$ at W1 = 17.2 mag (the CWISE 0738$-$6643 W1 magnitude). The median $\chi^2$ expected in this case (one degree of freedom) if the CatWISE motion uncertainties were correctly estimated would be 0.45, so the measured median $\chi^2$ of $\sim$2 indicates that the CatWISE motion uncertainties near the south ecliptic pole at W1 = 17.2 mag are typically underestimated by a factor of $\sqrt{2/0.45} = 2.1$.

According to the quoted CatWISE 2020 motion uncertainties, the CWISE 0738$-$6643 total proper motion is measured with a 1$\sigma$ fractional error of 3.5\%. Given that the CWISE 0738$-$6643 photometric distance uncertainty is at least $\sim$20\%, the proper motion contribution to the total $V_{tan}$ uncertainty is very subdominant, meaning that the total $V_{tan}$ uncertainty is relatively insensitive to the exact $\mu$ uncertainty adopted (for instance, inflating the Table \ref{tab:phot_cand} $\mu$ uncertainty for CWISE 0738$-$6643 by a factor of 2.1 would only increase the total $V_{tan}$ uncertainty by 5\%).

\subsection{WISEA J155349.96+693355.2}
\label{sec:w1553}

WISEA 1553+6933 \citep{byw_spitzer_phot} was first discovered by Backyard Worlds participant Nikolaj Stevnbak Andersen on 2017 December 9 using the WiseView image blinking tool\footnote{\url{http://byw.tools/wiseview-v2}}. This object was later independently rediscovered by Backyard Worlds citizen scientist David W. Martin. Despite its large proper motion ($\mu$ = 2157 $\pm$ 55 mas/yr), WISEA 1553+6933 was missed by prior \textit{WISE}-based brown dwarf searches, likely due to its faintness (W1 = 17.069 $\pm$ 0.034 mag, W2 = 15.615 $\pm$ 0.027 mag) and severe blending with a static background contaminant during the pre-hibernation $WISE$ mission phase in 2010 (see Figure \ref{fig:cutouts}, top row).

\begin{deluxetable}{lccc}
\tablecaption{WISEA J155349.96+693355.2 \label{tab:data}}
\tablehead{
\colhead{Parameter} & \colhead{Value} & \colhead{Ref.}}
\startdata
\cutinhead{Observed Properties}
$\mu$$_{\alpha}$ (mas yr$^{-1}$) & $-$1684 $\pm$ 56 & 1\\
$\mu$$_{\delta}$ (mas yr$^{-1}$)  & 1348 $\pm$ 53 & 1\\
$z_{AB}$ (mag) & 22.17 $\pm$ 0.21 & 3 \\
J$_{MKO}$ (mag) & 19.086 $\pm$ 0.065 & 2 \\
W1 (mag) & 17.069 $\pm$ 0.034 & 2\\
W2  (mag) & 15.615 $\pm$ 0.027 & 2\\
ch1 (mag) & 16.324 $\pm$ 0.019 & 1\\
ch2  (mag) & 15.458 $\pm$ 0.018 & 1\\
\cutinhead{Inferred Properties}
$T_{\rm eff}$ (K)  & 1200 $\pm$ 100 & 2\\
log($g$)  & $\approx$ 5.0-5.5 & 2\\
$[$Fe/H$]$ (dex) & $\approx$ $-$0.5 & 2\\
\enddata
\tablerefs{ (1) \citealt{byw_spitzer_phot} (2) this work (3) \citealt{nsc_dr1}. }
\end{deluxetable}

Table \ref{tab:data} lists the available photometric and astrometric properties for WISEA 1553+6933. Because WISEA 1553+6933 was blended during 2010, we opted to perform custom W1 and W2 photometry rather than use AllWISE \citep{cutri13} or CatWISE \citep{catwise_data_paper, catwise2020} photometry. Our custom photometry was obtained by running the \verb|crowdsource| pipeline \citep{decaps, unwise_catalog} on a set of coadded unWISE images \citep{tr_neo2}, each of which stacks together a single \textit{WISE} sky pass worth of exposures. There are 12 such coadded epochs available per band spanning 2014 to 2019, and the results of averaging the per sky pass W1 and W2 \verb|crowdsource| photometry are provided in Table \ref{tab:data}. The \textit{Spitzer} ch1 (3.6~$\mu$m) and ch2 (4.5~$\mu$m) photometry in Table \ref{tab:data} is from PID14076 \citep[PI: Faherty;][]{byw_spitzer_phot}. WISEA 1553+6933 was also detected at low significance by the Mayall $z$-band Legacy Survey \citep[MzLS;][]{dey_overview} with $z_{AB} = 22.17 \pm 0.21$ mag \citep{nsc_dr1}, though no brown dwarf absolute magnitude or color-type relations incorporating Mayall/Mosaic3 \citep{mosaic3} $z$-band currently exist, limiting the utility of this $z$-band data point.

WISEA 1553+6933 is undetected by 2MASS and too far north for UKIRT imaging. \cite{byw_spitzer_phot} obtained a 5$\sigma$ MKO J band limit of J $ > 17.34$ mag using the CPAPIR imager \citep{cpapir} at the Mont Megantic Observatory \citep{omm}. The corresponding J$-$W2 $>$ 1.73 mag color limit is consistent with either a T dwarf or T subdwarf scenario. We therefore sought deeper J band imaging to better constrain the nature of WISEA 1553+6933 and assess the prospects for near-infrared spectroscopy. On 2018 March 29, we obtained 25 $\times$ 60 s MKO J band exposures (PI: Wisniewski) of WISEA 1553+6933 with the NIRI instrument \citep{niri} on the 8.1 meter Gemini North telescope. The J band delivered image quality was 0.6$''$ and conditions were photometric. We reduced the Gemini J band images with IRAF\footnote{\url{http://ast.noao.edu/sites/default/files/A_Guide_to_Reducing_Near-IR_Images_1.pdf}} and calibrated the photometric zeropoint to 2MASS after converting 2MASS calibrator J magnitudes to the MKO system. The uncertainty on our Gemini J band magnitude in Table \ref{tab:data} is dominated by the tie-down to 2MASS, as a limited number of nearby 2MASS calibrators were available and these have only modest S/N in 2MASS. WISEA 1553+6933 is 1.54 (1.75) magnitudes  redder in J$-$W2 (J$-$ch2) color than would be expected for a typical T dwarf of its same W1$-$W2 (ch1$-$ch2) color. 

 The \textit{Spitzer}-based photometric distance estimate of $38.6^{+4.6}_{-4.2}$ pc for WISEA 1553+6933 from \cite{byw_spitzer_phot} implies an extremely high tangential velocity of $V_{tan} = 395^{+48}_{-44}$ km/s. The WISEA 1553+6933 phototype based on \textit{Spitzer} ch1-ch2 color is T5 $\pm$ 1 \citep{byw_spitzer_phot} and the polynomial relations from Table 13 of \cite{cw_byw_20pc} yield a \textit{Spitzer}-based temperature estimate of 1076 $\pm$ 89 K. Note that the \textit{Spitzer}-based photometric distance and temperature relations are fit to `normal' T dwarfs. If WISEA 1553+6933 is sub-luminous at ch2 relative to typical T dwarfs of the same ch1$-$ch2 color, its distance and tangential velocity will both be lower than the estimates reported here. It would be surprising if the WISEA 1553+6933 $V_{tan}$ were truly so high as $\sim$400 km/s, although it is the case that our motion selection methodology biases us toward finding objects with preferentially high $V_{tan}$. For example, the parent \textit{Spitzer} sample from which WISEA 1553+6933 is drawn \citep{ byw_spitzer_phot} has a median estimated $V_{tan}$ of 60 km/s, which is $\sim$2$\times$ higher than the median $V_{tan}$ of late  T and Y dwarfs within 20 pc from \cite{davy_parallaxes}. WISEA 1553+6933 has $H_{W2} = 22.28 \pm 0.06$ mag, a larger W2 reduced proper motion than that of any spectroscopically confirmed L or T subdwarf from the prior literature\footnote{Note, however, that WISEA 1553+6933 does not have the largest W2 reduced proper motion among all known brown dwarfs. For instance, the Y dwarf WISE 0855$-$0714 \citep{j0855} has $H_{W2} \approx 23.6$ mag.}, by a margin of $\sim$1.2 magnitudes (see Figure \ref{fig:sd_rpm}).

\subsection{CWISE J221706.28$-$145437.6}
\label{sec:w2217}

CWISE 2217$-$1454 was initially discovered by Dan Caselden using supervised machine learning methods to identify fast-moving, red objects based on proper motions and W1$-$W2 colors drawn from the CatWISE 2020 catalog (see $\S$3 of \citealt{marocco2019} for details of this selection methodology). CWISE 2217$-$1454 was also independently discovered by Backyard Worlds citizen scientists L\'eopold Gramaize, Sam Goodman, and Arttu Sainio.

CWISE 2217$-$1454 has W1$-$W2 = 1.65 $\pm$ 0.10 mag and is undetected in J band imaging from VHS. To probe deeper in the near-infrared, we obtained follow-up Keck/MOSFIRE \citep{mosfire} J band imaging on the night of 2020 September 3 (PI: Marocco). Observing conditions were excellent, with clear sky and $\sim$0.5$''$ seeing. We acquired 18 $\times$ 100 s  frames which were subsequently coadded and photometrically calibrated to 2MASS after converting the 2MASS calibrator J magnitudes to the MKO system. We find a high significance detection of CWISE 2217$-$1454: J = $20.66 \pm 0.02$ mag. With J$-$W2 = $4.88 \pm 0.06$ mag,  CWISE 2217$-$1454 is thus 2.93 magnitudes redder in J$-$W2 than would be expected for a typical T dwarf of its same W1$-$W2 color \citep{cw_byw_20pc}. Using VHS imaging we find a 5$\sigma$ limit of $K_S > 18.20$ mag, but this limit is not deep enough to materially constrain the nature of CWISE 2217$-$1454.

Photometric and astrometric properties of CWISE 2217$-$1454 are listed in Table \ref{tab:phot_cand}. In addition to its anomalously red J$-$W2 color, CWISE 2217$-$1454 has a very large total proper motion of $\mu = 1878 \pm 65$ mas/yr. Its W2 reduced proper motion of $H_{W2} = 22.14 \pm 0.09$ mag is larger than that of any spectroscopically confirmed L or T subdwarf from the prior literature, and only $\sim$0.15 mag lower than that of WISEA 1553+6933 (see Figure \ref{fig:sd_rpm}).

Using the \cite{cw_byw_20pc} polynomial relations fit to the bulk T dwarf population, we can derive \textit{WISE}-based phototype and distance estimates for CWISE 2217$-$1454. From the W1$-$W2 color, we obtain a spectral type estimate of T5.5 $\pm$ 1.2 and a photometric distance estimate of $39.4^{+9.5}_{-7.6}$ pc (based on $M_{W2}$ inferred from the $WISE$ color). This photometric distance estimate yields a corresponding $V_{tan}$ estimate of $351^{+85}_{-69}$ km/s. As for CWISE 0738$-$6643, if CWISE 2217$-$1454 is sub-luminous at WISE wavelengths relative to typical T dwarfs of the same W1$-$W2 color, its distance and tangential velocity will both be smaller than the present estimates. The CWISE 2217$-$1454 proper motion uncertainties quoted in Table \ref{tab:phot_cand} may be underestimated by $\sim$7\% based on Figure 13 of \cite{catwise2020}, which provides the most applicable assessment of the quoted CatWISE 2020 motion uncertainties for CWISE 2217$-$1454, given that CWISE 2217$-$1454 is near the ecliptic plane rather than the south ecliptic pole. Inflating the CWISE 2217$-$1454 proper motion uncertainties by $\sim$7\% would negligibly increase our quoted total $V_{tan}$ uncertainty, as the latter is strongly dominated by the large $\sim$20\% uncertainty on our photometric distance estimate.

\section{Spectroscopy}
\label{sec:spectro}

\begin{figure*}
\plotone{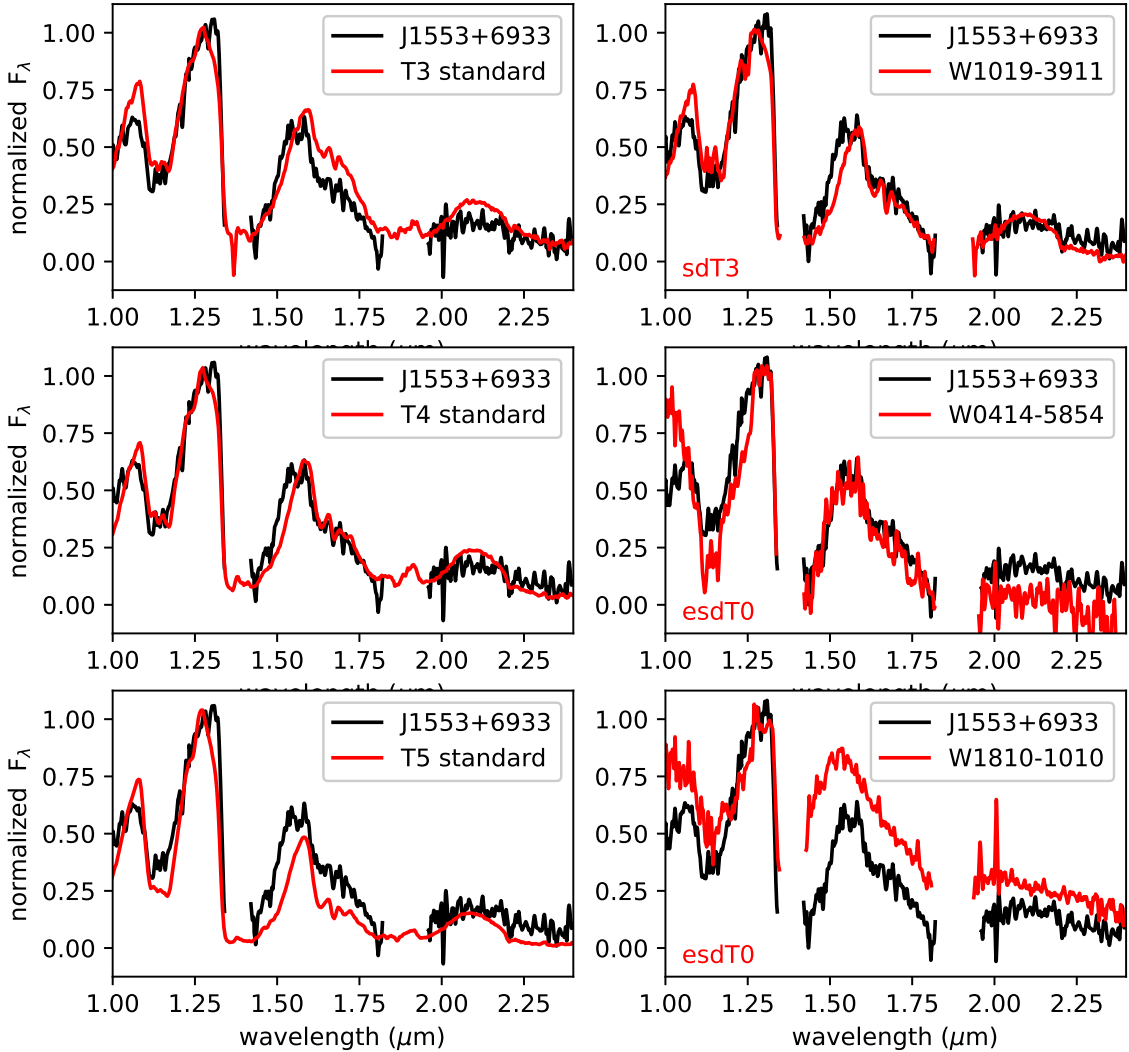}
\caption{Comparison of the WISEA 1553+6933 NIRES spectrum against spectral standards drawn from the SpeX Prism Library (data from \citealt{Burgasser04}; left column)
and a selection of T type subdwarfs (right column). The T4 standard provides the best visual match among field standards, although  relative to the standard WISEA 1553+6933 appears to have its H band peak shifted blueward, be slightly suppressed at K and display an excess on the red side of the J band peak. Top right: comparison against the sdT3 WISEA J101944.62$-$391151.6 \citep{schneider_neowise, greco}. Relative to the T3 standard, this sdT3 better captures the WISEA 1553+6933 spectrum's suppression at K band. Middle right: comparison against the esdT0 WISEA J041451.67$-$585456.7 \citep{esdTs}. Notably, of the comparison spectra shown, only WISEA J041451.67$-$585456.7 matches WISEA 1553+6933 in terms of the blue side of the H band peak. Bottom right: comparison against the esdT0 WISEA J181006.18$-$101000.5 \citep{esdTs}. All spectra are normalized to unity between 1.27~$\mu$m and 1.29~$\mu$m.}
\label{fig:standards_sdTs}
\end{figure*}

We used the Near-Infrared Echellette Spectrometer (NIRES; \citealt{Wilson2004}) on the Keck II telescope on 2020 July 7 (UT) to obtain 0.94-2.45~$\mu$m near infrared spectra of WISEA 1553+6933. Conditions were clear with 0$\farcs$9 seeing. Keck/NIRES has a fixed instrument configuration with a 0$\farcs$55 slit producing resolution $\sim$ 2,700 data. The target was visible in the $K$-band slit-viewing camera and placed into the spectroscopic slit. We obtained a set of 12 $\times$ 300 second spectroscopic frames obtained in an ABBA nodding pattern along the slit over an airmass range of 2.0--2.4. 
The A0~V star TYC 4282-488-1 ($V$ = 10.83 mag) was observed immediately afterward for flux calibration and telluric correction, and flat field lamp exposures were obtained for pixel response calibration.
Data were reduced using a modified version of Spextool (\citealt{Cushing04}; see also $\S$4.4 of \citealt{kirkpatrick11}), following the standard procedure which includes pixel calibration, wavelength calibration, and spatial and spectral rectification using flat field and telluric line exposures; optimal extraction of individual spectra from A-B pairwise subtracted frames; combination of these spectra with outlier masking; and telluric correction using the A0~V spectrum following \citet{vacca03}. The reduced, smoothed Keck/NIRES spectrum is shown in Figures \ref{fig:standards_sdTs}-\ref{fig:LOWZ_fits} in comparison to various standards, literature subdwarfs and atmospheric models.

The left column of Figure \ref{fig:standards_sdTs} compares the WISEA 1553+6933 spectrum against T3 (2MASS J12095613$-$1004008), T4 (2MASSI J2254188+312349) and T5 (2MASS J15031961+2525196) standards from the SpeX Prism Library \citep{burgasser06,SPL}. All spectra in Figure \ref{fig:standards_sdTs} are normalized to unity between 1.27~$\mu$m and 1.29~$\mu$m. The T4 standard provides the best visual match: the WISEA 1553+6933 spectrum at wavelengths of 1-1.3~$\mu$m is reasonably well replicated by the T4 standard. In the H band, WISEA 1553+6933 appears broader than the T4 standard, with the observed H band peak extending bluer than that of the T4 standard. Relative to the T4 standard, neither the T3 nor T5 standard better matches WISEA 1553+6933 at H band; the T3 standard is too elevated and broad in H band, while the T5 standard has an insufficiently strong, overly narrow H band peak. At K band, the T4 again provides the best match among the standards, although the WISEA 1553+6933 spectrum appears `flatter' than all three T standards in the 2-2.4~$\mu$m wavelength range. Such K band suppression is expected for T type subdwarfs due to collision-induced H$_2$ absorption, which becomes enhanced when metallicity is low and surface gravity is high \citep[e.g.,][]{saumon12}. We also note that, relative to the T standards, WISEA J1553+6933 displays a flux excess on the red side of its J band peak, a hallmark feature in the spectra of T subdwarfs \citep{wolf1130}. This subdwarf signature in the WISEA 1553+6933 spectrum is not well-replicated by any of our best-fitting models (Figures \ref{fig:phoenix_fits} and \ref{fig:LOWZ_fits}). Our finding that, among the standards, T4 best matches WISEA 1553+6933 is consistent with its \textit{Spitzer}-based phototype of T5 $\pm$ 1 from \cite{byw_spitzer_phot}.

Figure \ref{fig:standards_sdTs} also provides a comparison against the sdT3 WISEA J101944.62$-$391151.6 \citep[][hereafter WISEA 1019$-$3911]{schneider_neowise, greco}. At K band, the WISEA 1553+6933 spectrum's amplitude and shape are slightly better matched by WISEA 1019$-$3911 than by any of the T dwarf standards. Like the T standards, WISEA 1019$-$3911 also fails to match the H band peak shape observed for WISEA 1553+6933. Interestingly, the esdT WISEA 0414$-$5854 \citep{esdTs} matches WISEA 1553+6933 better at H band than either the sdT WISEA 1019$-$3911 or the T standards. The other esdT plotted \citep[WISEA 1810$-$1010;][]{esdTs} is dramatically enhanced at H band in a manner that does not match WISEA 1553+6933. The esdT spectra also exhibit major differences relative to WISEA 1553+6933 blueward of 1.3~$\mu$m. In particular, the \mbox{esdTs} show a strong flux enhancement at $\sim$1-1.1~$\mu$m not present in the WISEA 1553+6933 spectrum. The esdT K band spectra are poor matches to WISEA 1553+6933 --- WISEA 0414$-$5854 is much suppressed whereas WISEA 1810$-$1010 is significantly enhanced.

In $\S$\ref{sec:phoenix} and $\S$\ref{sec:lowz} we fit the WISEA 1553+6933 NIRES spectrum with two grids of low-metallicity model atmosphere spectra in order to quantify the metallicity and effective temperature of WISEA 1553+6933.

\subsection{PHOENIX Models}
\label{sec:phoenix}

\begin{figure*}
\plotone{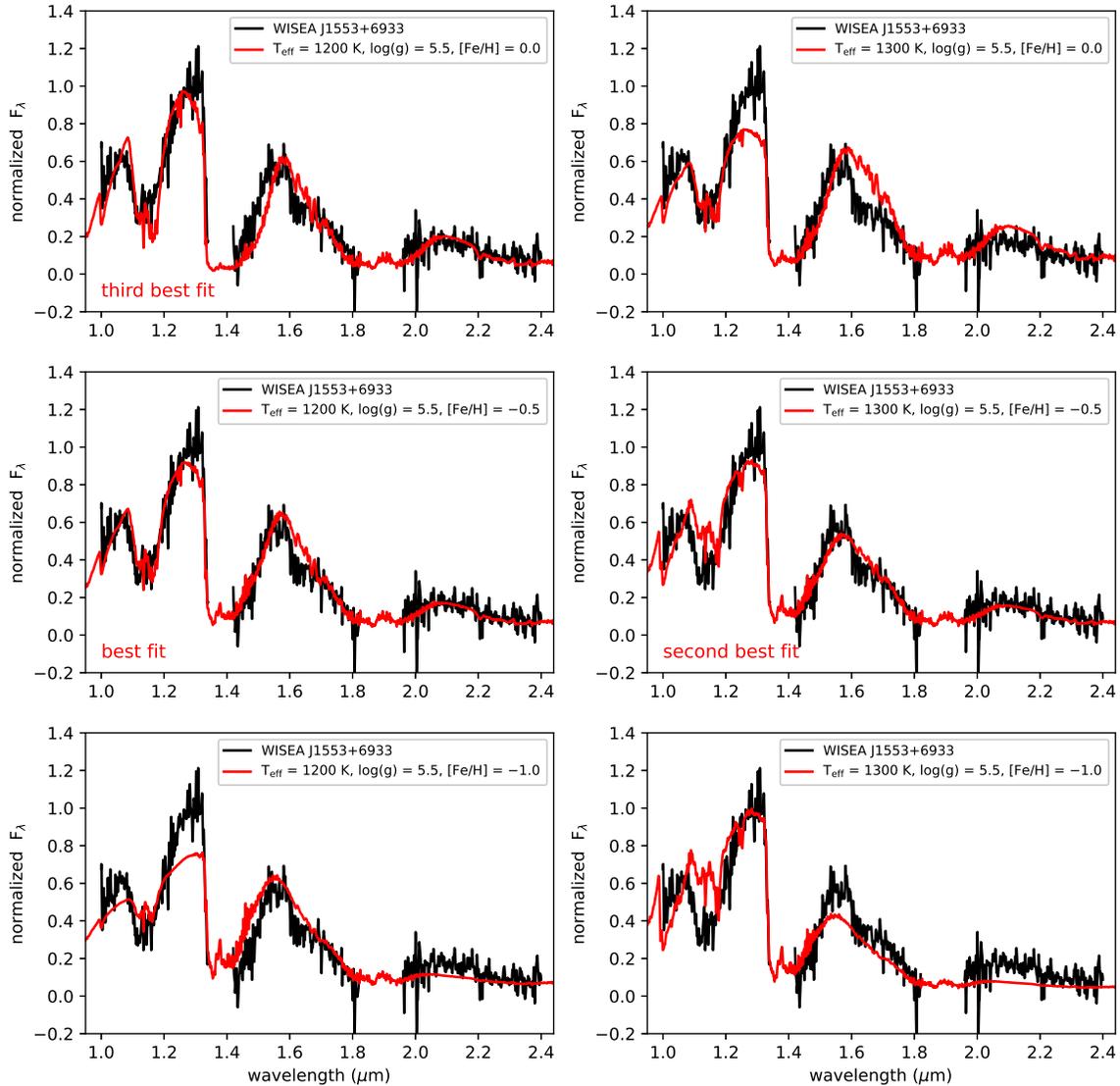}
\caption{Comparison of the WISEA 1553+6933 NIRES spectrum against various models from a PHOENIX model atmosphere grid extending to low-metallicity \citep{gerasimov_phoenix}. The top row's models have solar metallicity, the middle row's models have [Fe/H] = $-0.5$, and the bottom row's models have [Fe/H] = $-1.0$. The left column's models have $T_{\rm eff} = 1200$ K while the right column's models have $T_{\rm eff} = 1300$ K. The best-fit model has [Fe/H] = $-0.5$ and $T_{\rm eff} = 1200$ K. The second (third) best-fit model has [Fe/H] = $-0.5$ ([Fe/H] = 0) and $T_{\rm eff} = 1300$ K ($T_{\rm eff} = 1200$ K).}
\label{fig:phoenix_fits}
\end{figure*}

We compared the WISEA 1553+6933 spectrum against a set of PHOENIX atmospheric models \citep{phoenix99, phoenix13} extending to low metallicity\footnote{\url{http://atmos.ucsd.edu/?p=atlas}} \citep{gerasimov_phoenix}. In total, this PHOENIX model grid contains 44 spectra ranging in metallicity from [Fe/H] = 0 to [Fe/H] = $-3$ dex and from 900 K to 1500 K in effective temperature, with two log($g$) values available (5.0 and 5.5; $g$ in cgs units). Each model spectrum covers the 0.4-2.6~$\mu$m wavelength range. The log($g$) = 5.0 PHOENIX models all provide very poor fits to the WISEA 1553+6933 spectrum. This PHOENIX grid is rather sparse; for instance, the grid's native metallicity spacing is 1 dex. Overlaying the log($g$) = 5.5 models at $T_{\rm eff} = 1200$-1300 K as in Figure \ref{fig:phoenix_fits}, it immediately became clear that this would be the best-fit temperature range for WISEA 1553+6933 and that the optimal metallicity would lie somewhere between [Fe/H] = $-1$ and [Fe/H] = 0. We therefore interpolated the PHOENIX grid in metallicity to provide model templates at half dex intervals. This interpolation is one-dimensional in the sense that we only interpolate between a pair of models separated by 1 dex in [Fe/H] when those two models are identical in all other parameters, including $T_{\rm eff}$ and log($g$). This interpolation procedure results in 25 additional models with half-integer metallicity values of [Fe/H] = $-0.5$, $-1.5$, and $-2.5$ dex. As an example, in the left column of Figure \ref{fig:phoenix_fits}, the middle panel's model ([Fe/H] = $-0.5$, $T_{\rm eff} = 1200$ K, log($g$) = 5.5) is an interpolation in metallicity between the model shown in the top panel  of that column ([Fe/H] = 0, $T_{\rm eff} = 1200$ K, log($g$) = 5.5) and the model shown in the bottom panel of that column ([Fe/H] = $-1$, $T_{\rm eff} = 1200$ K, log($g$) = 5.5). The same applies to the right column of Figure \ref{fig:phoenix_fits}, except for the case of $T_{\rm eff} = 1300$ K.

We fit all 69 models in our augmented PHOENIX grid to the observed WISEA 1553+6933 spectrum, using the $\chi^2$ goodness of fit metric. When fitting, the observed spectrum is normalized to unity at the J band peak, and the overall normalization of each model (across the full 1-2.4~$\mu$m wavelength range) is treated as a single free parameter. Figure \ref{fig:phoenix_fits} shows a comparison of the WISEA 1553+6933 spectrum against six PHOENIX models spanning the reasonably well-fitting range of temperature and metallicity: $1200 \ \rm{K} \le \it{T}_{\rm eff} \le \rm{1300}$ K, $-1 \le $ [Fe/H] $ \le 0$ and log($g$) = 5.5. The three best-fitting models are labeled in Figure \ref{fig:phoenix_fits}. The best-fit model has $T_{\rm eff}$ = 1200 K, [Fe/H] = $-0.5$ dex and log($g$)= 5.5 (left column, middle row of Figure \ref{fig:phoenix_fits}). Among the set of three best-fitting models, two have [Fe/H] = $-0.5$, one has [Fe/H] = 0, and all have $T_{\rm eff} = $ 1200 K or 1300 K.

The $T_{\rm eff} = 1200$ K, [Fe/H] = 0 model mostly matches the WISEA 1553+6933 spectrum, but like the T4 standard ($\S$\ref{sec:spectro}) this model fails to capture the observed H band peak's blue extension. The $T_{\rm eff} = 1200$ K, [Fe/H] = $-1$ model has a broadened H band peak, but this model's H band peak is much stronger relative to J and K than is observed for WISEA 1553+6933. The best-fit model ($T_{\rm eff} = 1200$ K, [Fe/H] = $-0.5$) accurately matches the blue wing of the H band peak while also displaying the correct relative amplitudes for all three spectral segments between 1.0 and 2.4~$\mu$m. For $T_{\rm eff}$ = 1300 K, the observed H band peak is again best reproduced with [Fe/H] = $-0.5$, but the  agreement at 1-1.3~$\mu$m is substantially worse throughout the $-1 \le$ [Fe/H] $\le 0$ metallicity range.

\subsection{New LOWZ Models}
\label{sec:lowz}

We have also generated an extensive new grid of low-temperature, low-metallicity model atmosphere spectra and compared these against the near-infrared spectrum and broadband colors of WISEA 1553+6933. Since we expect this new model grid, which we refer to as `LOWZ', to be generally useful for studies of brown dwarfs, we have made these models publicly available online\footnote{\url{https://doi.org/10.7910/DVN/SJRXUO}}.


We use the ScCHIMERA 1D-radiative-convective-thermochemical equilibrium tool \citep{piskorz18, gharib19, arcangeli18, mansfield18, zalesky19, baxter20, beatty20, colon20} to compute our cloud-free model grid.  Briefly, the grid solves for the net radiative fluxes using the \cite{toon89} two stream source function technique. The equilibrium temperature structure is achieved using a Newton-Raphson scheme \citep[e.g.,][]{mckay89}. Chemical equilibrium is computed using the NASA Chemical Equilibrium with Applications\footnote{\url{https://cearun.grc.nasa.gov/}} routine assuming rainout chemistry and scaled \cite{lodders09} solar abundances. We include a suite of relevant opacities (H$_2$-H$_2$/He collision-induced absorption \citep{richard12}, H$^{-}$ bound-free/free-free \citep{john88}, H$_2$O \citep{polyansky18}, CO \citep{rothman10}, CH$_4$ \citep{yurchenko14}, NH$_3$ \citep{yurchenko11}, H$_2$S \citep{tennyson12}, PH$_3$ \citep{phosphine}, HCN \citep{harris08}, C$_2$H$_2$ \citep{rothman13}, TiO \citep{mckemmish19}, VO \citep{mckemmish16}, SiO \citep{barton13}, CaH \citep{alavi18, yadin12}, MgH \citep{yadin12, gharib13}, CrH \citep{burrows05}, AlH \citep{yurchenko18}, FeH \citep{wende10},  Na, K \citep{allard07, allard19}), as well as several other atomics sourced from \cite{kurucz93} converted into R = 250 correlated-K coefficients (with 10 Gauss-quadrature points per bin, in double Gauss form) mixed on the fly using the \cite{amundsen17} resort-rebin procedure. Convection is treated as an additional flux via mixing length theory.  

\begin{figure*}
\plotone{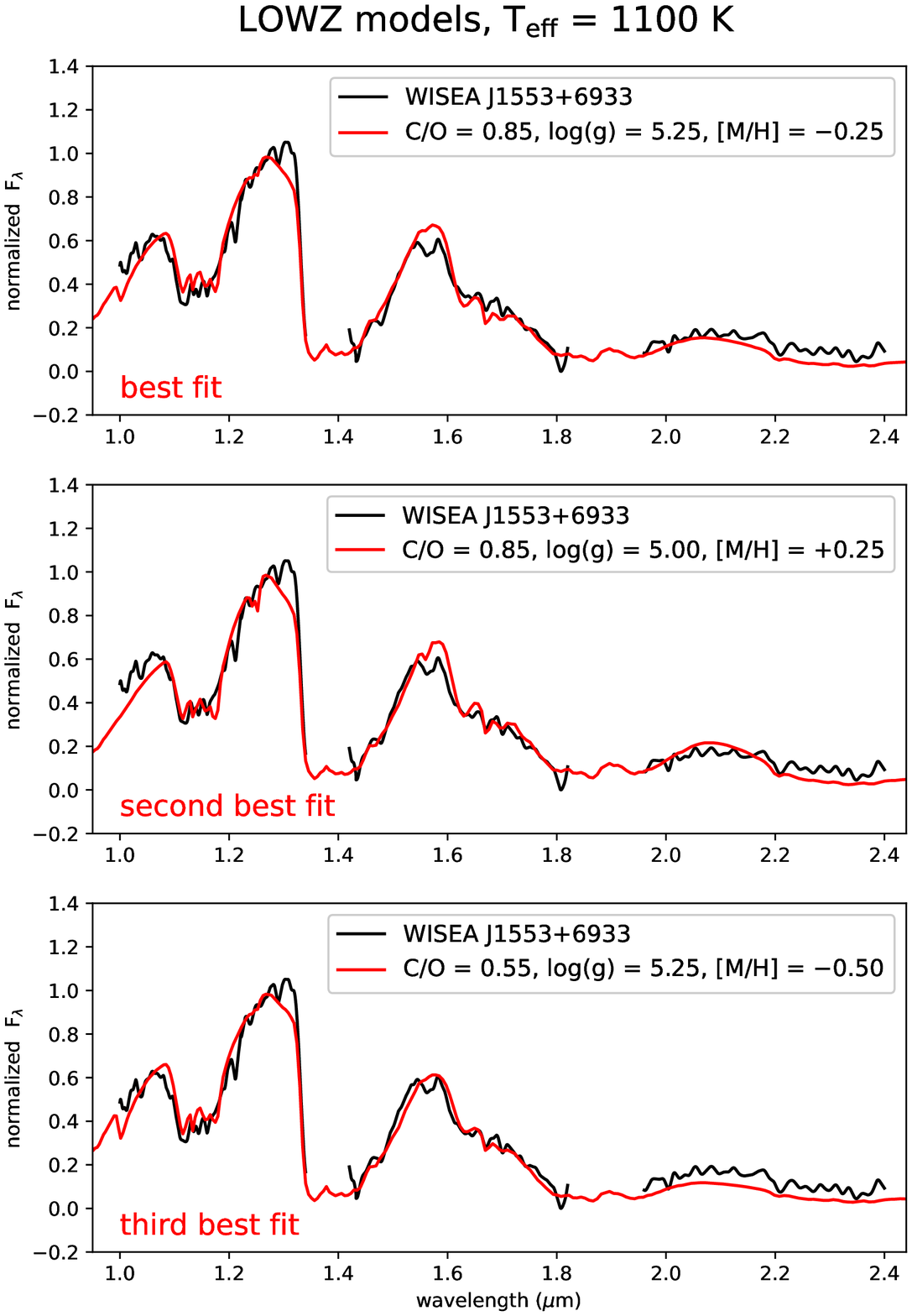}
\caption{Comparison of the WISEA 1553+6933 NIRES spectrum against the three best-fit models from our new LOWZ grid. All three best-fit models have $T_{\rm eff} = 1100$ K and log$_{10}(K_{zz})$ = 10. The best-fit LOWZ model has a slightly subsolar metallicity, with [M/H] = $-0.25$. The second best-fit LOWZ model has relatively low log($g$) = 5.0 and a slightly super-solar metallicity,  [M/H] = +0.25 dex. The third best-fit LOWZ model has [M/H] = $-0.5$.}
\label{fig:LOWZ_fits}
\end{figure*}

Each model in the LOWZ grid extends from 0.1-10~$\mu$m in wavelength coverage. The wavelength grid spacing is 4.6 nm at $\lambda = 1.15$~$\mu$m, 6.6 nm at $\lambda = 1.65$~$\mu$m, and 8.8 nm at $\lambda = 2.2$~$\mu$m. The LOWZ model metallicities range from [M/H] = $-$2.5 to +1.0 dex, sampling  values of [M/H] = $-2.5$, $-2$, $-1.5$, $-1$, $-0.5$, $-0.25$, 0, +0.25, +0.5, +0.75, +1 dex. The LOWZ effective temperatures span from $T_{\rm eff} = 500$ K to $T_{\rm eff} = 1600$ K, with values of $T_{\rm eff} $ =  500, 550, 600, 650, 700, 750, 800, 850, 900, 950, 1000, 1100, 1200, 1300, 1400, 1500, and 1600 K. The grid contains three distinct carbon-to-oxygen (C/O) ratio values (0.1, 0.55, 0.85), five log($g$) values (3.5, 4, 4.5, 5, 5.25), and three vertical eddy diffusion coefficient options, log$_{10}(K_{zz})$ = ($-$1, 2, 10), with $K_{zz}$ in units of cm$^2$/s. In all, our LOWZ grid contains 8,402 model atmosphere spectra sampling in the five-dimensional space of ([M/H], $T_{\rm eff}$, C/O, log($g$), log$_{10}(K_{zz})$).

We fit all 8,402 LOWZ models to the observed WISEA 1553+6933 spectrum, again using the $\chi^2$ goodness-of-fit metric. When fitting, the observed spectrum is normalized to unity at the J band peak, and the overall normalization of each model (across the full 1-2.4~$\mu$m wavelength range) is treated as a single free parameter. Figure \ref{fig:LOWZ_fits} shows the WISEA 1553+6933 NIRES spectrum overlaid on the three best-fitting LOWZ models. All three best-fitting models have $T_{\rm eff} = 1100$ K and log$_{10}(K_{zz})$ = 10. The best-fit LOWZ model has a slightly subsolar metallicity, with [M/H] = $-$0.25, C/O = 0.85 and log($g$) = 5.25. The second best-fit LOWZ model is slightly metal-enhanced with [M/H] = $+$0.25 and lower log($g$) = 5 rather than 5.25. The third best-fit LOWZ model has [M/H] = $-$0.5, log($g$) = 5.25 and a C/O ratio of 0.55.

One peculiarity of our best-fit LOWZ model's parameters is that the metallicity ([M/H] = $-0.25$ dex) is slightly subsolar, but the C/O value is super-solar. Intriguingly, this situation is reminiscent of  retrieval results for the T7.5pec dwarf SDSS J1416+1348B, a companion to the sdL7 SDSS J1416+1348A \citep{j1416, line_j1416}. \cite{j1416} found a slightly subsolar metallicity ([M/H] $\approx$ $-0.3$ dex) for the SDSS J1416+1348AB system, but a roughly solar C/O ratio. With $T_{\rm eff} \approx 600$-700 K \citep{j1416}, SDSS J1416+1348B is significantly colder than WISEA 1553+6933, so that it is not immediately evident whether the same spectral features are driving the elevated fits of C/O relative to [M/H] in both cases.

Taking into account the literature comparisons and model fits shown in Figures \ref{fig:standards_sdTs}-\ref{fig:LOWZ_fits}, we assign WISEA 1553+6933 a spectral type of sdT4 $\pm$ 0.5. The subtype is quite well-constrained given that the T3 and T5 standards in Figure \ref{fig:standards_sdTs} are very clearly worse matches to the WISEA 1553+6933 spectrum than is the T4 standard. We adopt a metallicity of $\approx -0.5$ dex based on the best-fit models shown in Figures \ref{fig:phoenix_fits} and \ref{fig:LOWZ_fits}. In Table \ref{tab:data} we quote the WISEA 1553+6933 effective temperature as $T_{\rm eff}$ = 1200 $\pm$ 100 K; the PHOENIX models favor temperatures of 1200-1300 K and the LOWZ models favor a temperature of 1100 K. The effective temperature from spectroscopic modeling is in good agreement with the photometrically estimated $T_{\rm eff} = 1076 \pm 89$ K value from $\S$\ref{sec:w1553}.

\section{LOWZ Model Synthetic Photometry}
\label{sec:synth}

\begin{figure*}
\plotone{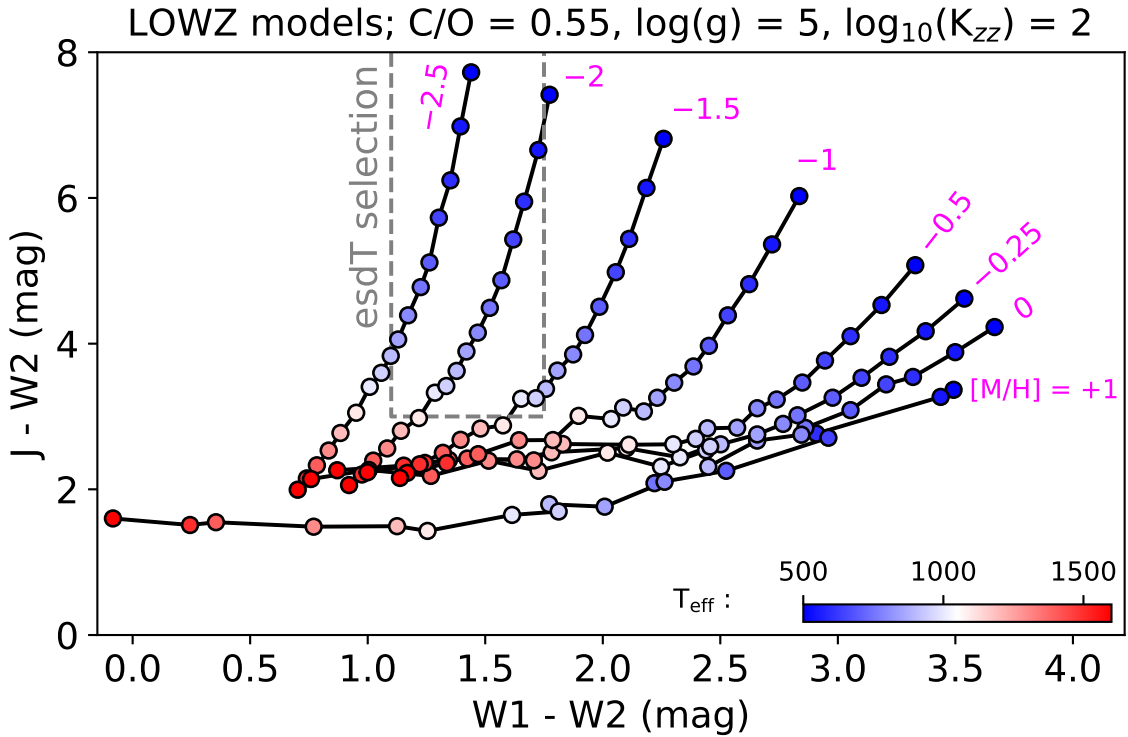}
\caption{LOWZ model synthetic colors for ``normal'' brown dwarf parameters of C/O = 0.55, log($g$) = 5.0 and log$_{10}$($K_{zz}$) = 2. Each connected set of data points represents the temperature sequence (500 K $\le T_{\rm eff} \le $ 1600 K) for one metallicity value. The metallicities range from super-solar ([M/H] = +1) to very metal poor ([M/H] = $-$2.5). Magenta annotations list each curve's [M/H] value. Each data point is color-coded by temperature, with blue being cold and red being relatively warm according to the inset color bar. The esdT candidate color-color selection box used in this work is shown as a dashed gray outline. As metallicity decreases from solar, the loci progressively compress and shift blueward in W1$-$W2 while expanding and shifting redward in J$-$W2, leading the [M/H] $\lesssim$ $-1.5$ curves to enter our esdT selection box.}
\label{fig:synth_color_color}
\end{figure*}

Because the LOWZ models extend from blueward of 1~$\mu$m all the way to 10~$\mu$m, we can generate synthetic near and mid-infrared (MIR) colors corresponding to each model's set of physical parameters. Using MKO, \textit{WISE} and \textit{Spitzer} transmission curves from the SVO Filter Profile Service \citep{svo_filters_1, svo_filters_2}, we computed each LOWZ model's J$-$W2, W1$-$W2, J$-$ch2 and ch1$-$ch2 colors in the Vega system. Figure \ref{fig:synth_color_color} shows our LOWZ model synthetic colors in the same J$-$W2 versus W1$-$W2 color-color space used to select this work's esdT candidates (see Figure \ref{fig:color_color}).

Figure \ref{fig:synth_color_color} includes synthetic colors for 136 LOWZ models, selected to have nominal ``normal'' brown dwarf parameters: C/O = 0.55, log($g$) = 5 and log$_{10}$($K_{zz}$) = 2. Each connected curve represents a unique [M/H] value's temperature sequence, from 500 K to 1600 K. The eight [M/H] curves shown span from [M/H] = +1 to [M/H] = $-$2.5. The [M/H] = 0 curve agrees reasonably well with the observed locus of ``normal'' T dwarfs shown as magenta points in Figure \ref{fig:color_color}. A dramatic trend can be seen among the set of temperature sequence curves as metallicity becomes progressively lower. As metallicity decreases from [M/H] = 0, the locus compresses and shifts blueward in terms of W1$-$W2, while expanding to push redward in J$-$W2. This metallicity evolution of the T dwarf sequence means that models with [M/H] $\lesssim$ $-$1.5 enter our esdT candidate color-color selection box from $\S$\ref{sec:candidates}, whereas the higher metallicity models (including super-solar) shown do not occupy this area. Thus, our LOWZ models predict the strong J$-$W2 color excess signature of the \cite{esdTs} esdTs. Our LOWZ model synthetic photometry also points to the possibility of finding very low metallicity T dwarfs with J$-$W2 excesses even larger than observed for any of the subdwarfs in this work or \cite{esdTs}, 5 $<$ J$-$W2 $<$ 8. This motivates future brown dwarf searches and follow-up campaigns to seek examples of such extreme J$-$W2 color at modest W1$-$W2 color in the $\sim$1$-$2 mag range.

\begin{figure}
\plotone{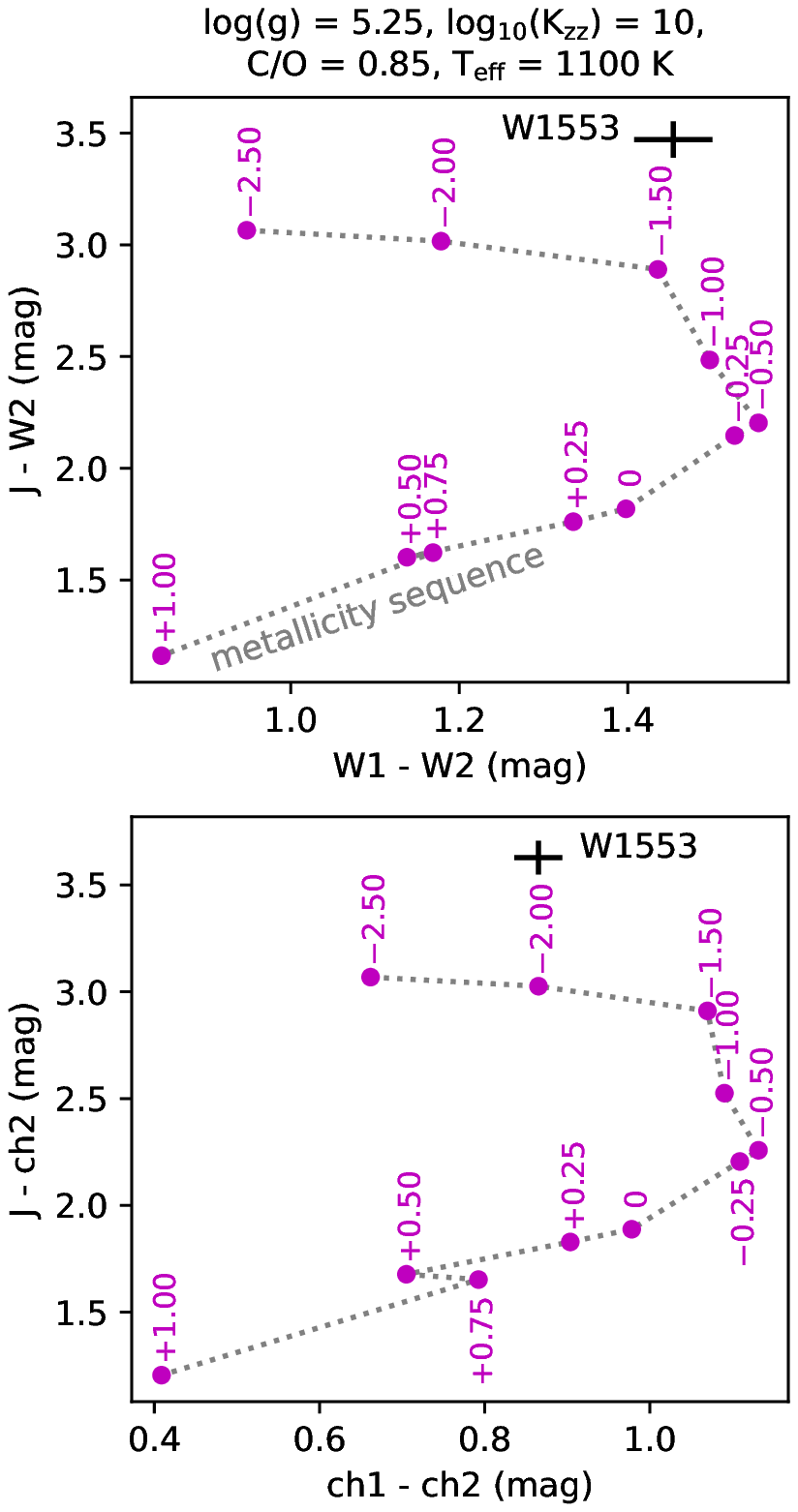}
\caption{Synthetic colors of our LOWZ models as a function of metallicity, fixing all other physical parameters to those of the best-fit WISEA 1553+6933 model from $\S$\ref{sec:lowz}. Magenta annotations provide the [M/H] value of each data point. WISEA 1553+6933 is shown as a black plus mark in each panel. Top: \textit{WISE-based} synthetic colors. Bottom: \textit{Spitzer}-based synthetic colors. The \textit{WISE}-based (\textit{Spitzer}-based) synthetic broadband colors favor a very low photometric metallicity of [M/H] = $-1.5$ ($-$2) for WISEA 1553+6933.}
\label{fig:photo_z}
\end{figure}

Returning to the case study of WISEA 1553+6933, we can also examine the metallicity sequence in this same synthetic color-color space while fixing all other physical parameters to be those of the best-fit LOWZ model from Figure \ref{fig:LOWZ_fits}. Figure \ref{fig:photo_z} shows the resulting metallicity sequence for log($g$) = 5.25, log$_{10}$($K_{zz}$) = 10, C/O = 0.85, and $T_{\rm eff}$ = 1100 K. The top panel illustrates the trend of J$-$W2 versus W1$-$W2, and the bottom panel shows the \textit{Spitzer}-based analog which replaces W1 (W2) with ch1 (ch2). WISEA 1553+6933 is closest to the [M/H] = $-1.5$ ($-2.0$) model prediction in the \textit{WISE}-based (\textit{Spitzer}-based) color-color diagram. Thus, synthetic broadband colors of the LOWZ models yield a `photometric metallicity' of $-1.5$ to $-2$ for WISEA 1553+6933, much lower than the best-fit [M/H] = $-0.25$ value derived from fitting the NIR spectral morphology ($\S$\ref{sec:lowz}). In Table \ref{tab:data} we retain the relatively high metallicity value derived from our $\S$\ref{sec:spectro} fits, since we favor the spectroscopic analysis over broadband photometric estimates. WISEA 1553+6933 illustrates that although the LOWZ models qualitatively capture observed trends of J$-$W2 versus W1$-$W2, there is tension between the metallicities implied by NIR spectral morphology as compared to NIR-MIR broadband colors. JWST low-resolution spectroscopy of objects like WISEA 0414$-$5854, WISEA 1810$-$1010 or WISEA 1553+6933 covering 1-5$\mu$m with NIRSpec \citep{jwst, jwst_nirspec} would provide valuable insight into the detailed successes and mismatches of our LOWZ models in this mid-infrared wavelength range, helping to understand the key physical processes/parameters shaping the spectra and colors of T subdwarfs.

\section{Conclusion}
\label{sec:conclusion}

We have presented discoveries and follow-up of three T type subdwarf candidates identified by the Backyard Worlds: Planet 9 citizen science project. These objects have large total proper motions in the range of 0.9-2.2$''$/yr, with correspondingly high tangential velocity estimates ($\sim$180-400 km/s) indicative of membership in the Milky Way thick disk or halo. Their photometry places them in the same unusual region of color-color space as the first two known examples of the esdT spectral class \citep{esdTs}. We have also published an extensive new grid of low temperature, low metallicity model atmospheres. Going forward, these models will aid in the characterization/classification of the T subdwarf population.

Keck/NIRES spectroscopy indicates that WISEA 1553+6933 is a T4 subdwarf with an effective temperature of $T_{\rm eff} = 1200 \pm 100$ K and metallicity $\approx -0.5$ dex. Although WISEA 1553+6933 has the largest W2 reduced proper motion among all known L and T type subdwarfs, suggesting that it may have unusually high kinematics and/or low luminosity even relative to other T subdwarfs, its near infrared spectral morphology does not indicate a metallicity sufficiently low to be classified as an esdT. On the other hand, comparison of the WISEA 1553+6933 broadband photometry from $\sim$1-5~$\mu$m against the LOWZ models favors a lower metallicity $\lesssim -1.5$ dex.

Our understanding of all three objects presented in this work would benefit from additional astrometry, in order to obtain trigonometric parallaxes. Trigonometric parallaxes would specify how sub-luminous these subdwarfs are relative to `normal' T dwarfs. Determining luminosities of the two known esdTs and other potentially similar objects may help to better map/define the sdT/esdT sequence.

Additional near-infrared photometry of this study's sample is also needed. In particular a J band detection of CWISE 0738$-$6643 would allow us to pinpoint its location in color-color space. Deeper H and K band follow-up imaging is needed for all three members of the present sample, and would allow us to place these objects within additional color-color diagrams (e.g., Figure 3 of \citealt{esdTs}).

Lastly, near-infrared spectroscopy is needed for the two new discoveries presented in this work, CWISE 0738$-$6643 and CWISE 2217$-$1454, to gauge their metallicities and temperatures, thereby determining whether either is a new member of the esdT spectral class.

Backyard Worlds will continue mining the $WISE$/NEOWISE data set for faint moving objects in search of more surprising substellar discoveries with anomalous properties.

\begin{center}
ACKNOWLEDGMENTS
\end{center}


The Backyard Worlds: Planet 9 team would like to thank the many Zooniverse volunteers who have participated in this project, from providing feedback during the beta review stage to classifying flipbooks to contributing to the discussions on TALK. We would also like to thank the Zooniverse web development team for their work creating and maintaining the Zooniverse platform and the Project Builder tools. This research was supported by NASA grant 2017-ADAP17-0067. This material is based upon work supported by the National Science Foundation under Grant No. 2007068, 2009136, and 2009177. F.M. acknowledges support from grant 80NSSC20K0452 under the NASA Astrophysics Data Analysis Program. SLC acknowledges the support of a STFC Ernest Rutherford Fellowship. The CatWISE effort is led by the Jet Propulsion Laboratory, California Institute of Technology, with funding from NASA's Astrophysics Data Analysis Program. Support for this work was provided by NASA through the NASA Hubble Fellowship grant \textit{HST}-HF2-51447.001-A awarded by the Space Telescope Science Institute, which is operated by the Association of Universities for Research in Astronomy, Inc., for NASA, under contract NAS5-26555. The work of PRME was carried out at the Jet Propulsion Laboratory, California Institute of Technology, under a contract with NASA. This publication makes use of data products from the \textit{Wide-field Infrared Survey Explorer}, which is a joint project of the University of California, Los Angeles, and the Jet Propulsion Laboratory/California Institute of Technology, funded by the National Aeronautics and Space Administration. The UHS is a partnership between the UK STFC, The University of Hawaii, The University of Arizona, Lockheed Martin and NASA. The VISTA Data Flow System pipeline processing and science archive are described in \cite{irwin04}, \cite{hambly08} and \cite{vsa}. We have used data from VHS DR6. This research has made use of the NASA/IPAC Infrared Science Archive, which is funded by the National Aeronautics and Space Administration and operated by the California Institute of Technology. Some of the data presented herein were obtained at the W. M. Keck Observatory, which is operated as a scientific partnership among the California Institute of Technology, the University of California and the National Aeronautics and Space Administration. The Observatory was made possible by the generous financial support of the W. M. Keck Foundation. The authors wish to recognize and acknowledge the very significant cultural role and reverence that the summit of Maunakea has always had within the indigenous Hawaiian community.  We are most fortunate to have the opportunity to conduct observations from this mountain. Based on observations obtained at the international Gemini Observatory, a program of NSF's NOIRLab, which is managed by the Association of Universities for Research in Astronomy (AURA) under a cooperative agreement with the National Science Foundation on behalf of the Gemini Observatory partnership: the National Science Foundation (United States), National Research Council (Canada), Agencia Nacional de Investigacion y Desarrollo (Chile), Ministerio de Ciencia, Tecnologia e Innovacion (Argentina), Ministerio da Ciencia, Tecnologia, Inovacoes e Comunicacoes (Brazil), and Korea Astronomy and Space Science Institute (Republic of Korea). This research has made use of the VizieR catalogue access tool, CDS, Strasbourg, France (DOI : 10.26093/cds/vizier). The original description of the VizieR service was published in \cite{vizier}. This research has made use of the SVO Filter Profile Service (http://svo2.cab.inta-csic.es/theory/fps/) supported from the Spanish MINECO through grant AYA2017-84089.

\vspace{5mm}
\facilities{Keck(NIRES,~MOSFIRE), \textit{WISE}, \textit{Spitzer}(IRAC), Gemini North(NIRI)}

\software{WiseView \citep{wiseview}, IRAF \citep{iraf}, SPLAT \citep{SPLAT}}

\bibliography{ms}{}
\bibliographystyle{aasjournal}

\end{document}